\documentclass[%
 aip,
 amsmath,amssymb,
 reprint,%
]{revtex4-1}

\usepackage{graphicx}
\usepackage{dcolumn}
\usepackage{bm}
\usepackage{xcolor}

\usepackage[utf8]{inputenc}
\usepackage[T1]{fontenc}
\usepackage{mathptmx}

\begin{document}

\preprint{AIP/123-QED}

\title[]{Markov-modulated model for landing flow dynamics: An ordinal analysis validation}

\author{F. Olivares}
\email{felipe@ifisc.uib-csic.es}
\affiliation{Instituto de F\'isica Interdisciplinar y Sistemas Complejos IFISC (CSIC-UIB), Campus UIB, Palma de Mallorca, 07122, Spain.}
\author{L. Zunino}%
\affiliation{Centro de Investigaciones \'Opticas (CONICET La Plata-CIC-UNLP), 1897 Gonnet, La Plata, Argentina,}
\affiliation{Departamento de Ciencias B\'asicas, Facultad de Ingenier\'ia, Universidad Nacional de La Plata (UNLP), 1900 La Plata, Argentina.}
\author{M. Zanin}
\affiliation{Instituto de F\'isica Interdisciplinar y Sistemas Complejos IFISC (CSIC-UIB), Campus
UIB, Palma de Mallorca, 07122, Spain.}

\date{\today}

\begin{abstract}
Air transportation is a complex system characterised by a plethora of interactions at multiple temporal and spatial scales; as a consequence, even simple dynamics like sequencing aircraft for landing can lead to the appearance of emergent behaviours, which are both difficult to control and detrimental to operational efficiency. We propose a model, based on a modulated Markov jitter, to represent ordinal pattern properties of real landing operations in European airports. The parameters of the model are tuned by minimising the distance between the probability distributions of ordinal patterns generated by the real and synthetic sequences, as estimated by the Permutation Jensen-Shannon Distance. We show that the correlation between consecutive hours in the landing flow changes between airports, and that it can be interpreted as a metric of efficiency. We further compare the dynamics pre and post COVID-19, showing how this has changed beyond what can be attributed to a simple reduction of traffic. We finally draw some operational conclusions, and discuss the applicability of these findings in a real operational environment.
\end{abstract}

\maketitle

\begin{quotation}
Air traffic flows compose a complex dynamics, result of the multiple interactions that may emerge between aircraft. To illustrate, airports have a limited landing capacity; the late arrival of one aircraft in a saturated airport may force other aircraft to delay their respective  landings, creating a knock-off effect. Identifying such chains of events is not always simple; yet, their characterisation, and especially the quantification of the temporal scale over which such interactions propagate, is a key element for evaluating the performance of operations. If directly identifying interactions is not an option, a solution is to analyse the fingerprint they leave in time series representing the aggregated dynamics of the system---in this case, in time series representing hourly landing volumes. We here use a simple dynamical model (specifically, a modulated Markov jitter) to represent such dynamics, and tune it implementing a recently introduced ordinal approach on real data. The result is used to extract a mean life-time, representing, for each airport, how long can interactions between landings propagate. In more general terms, this contribution illustrates yet another example of how statistical physics concepts can be used to obtain insights about real-world technical systems, even when only macro-scale information is available.
\end{quotation}

\section{\label{sec:intro}Introduction}

Studying the dynamics of complex systems for which we have limited information is a major challenge, which researchers frequently have to face when dealing with many real-world problems. If proper knowledge about the underlying structure and dynamics is not available, and thus a micro-scale model cannot be reconstructed, the natural solution is to resort to the study of macro-scale observable, \textit{i.e.} to a data-driven approach. This has lead, in recent years, to the widespread adoption of Machine Learning and Deep Learning in science \cite{goh2017deep, ching2018opportunities, tanaka2021deep}---in spite of some major drawbacks, including their black-box nature, and hence the lack of interpretability \cite{li2022interpretable}. Mixed approaches, combining simple principled models and data tuning, are nevertheless not new in physics, and can be traced back to the first models of our solar system \cite{murschel1995structure}. They present the advantage of being easily interpretable, and thus enabling the analysis of specific aspects of the system; but also of being simplified representations tuned according to real data, and hence not requiring a complete knowledge.

Within this vast field, of special interest is the creation of models able to reproduce some characteristics of a dynamical process (\textit{i.e.} a time series) based on a limited set of parameters. These parameters can be fitted against real data, in order to understand the underlying properties of the real system; but can further be modified to simulate how the system would react to internal or external perturbations. Examples of this, of relevance for the present study, include the use of modulated short-range correlated noise to represent the daily sampled sunspot number sequence~\cite{blanter2005solar}, and temperature fluctuation in air avalanches~\cite{crouzeix2006long}. More concretely, these models are based on a high-frequency Markov process, multiplied by a long-term component defining the modulated Markov jitter~\cite{blanter2005solar}. 

A system that has mostly been neglected by the physical community is air transport, and specifically the problem of understanding the mechanisms governing air traffic flows. In spite of being a highly engineered and regulated system, and thus being completely man-made, it is also defined by the interactions among numerous elements, and thus by a highly complex dynamics \cite{cook2015applying}; as a result, unexpected (emergent) patterns frequently appear. To illustrate, airport and airspaces have a limited capacity, constraining the number of aircraft that can operate within them in a given time window. In a saturated context, the late arrival of one aircraft (\textit{i.e.} its arrival in a time window that is not the originally planned one) may result in another aircraft to be delayed, and in a knock-off (or avalanche) effect.
The study of this problem has mainly been tackled through mathematical models~\cite{bayen2006adjoint,zhang2015analysis,zhang2017impacts,menon2006computer,yang2017fundamental,wei2013total}, which are nevertheless limited by the lack of complete knowledge about the underlying interactions. On the other hand, macroscopic analyses, \textit{e.g.} based on the characterisation of real time series from the view point of statistical physics, are scant at best. Some examples include the study of landing volume sequences through multifractal analysis~\cite{zhang2019multifractal,liu2020multiscale}, visibility graphs~\cite{liu2018exploring,li2020exploring}, or the study of the irreversibility of delay time series~\cite{zanin2021assessing}.

In this contribution we propose the use of a Markov-modulated model to represent the ordinal pattern properties of the low and high-frequency dynamical components of hourly landing flows. More specifically, we consider two dynamical components: a high-frequency term, given by an auto-regressive process; and a low-frequency one, simulating the inactivity during nights and the reduced activity around mid-day (as observed in real data). The parameters of this model are tuned to match the ordinal dynamics observed in real data, by minimising the distance between these and the generated time series; such distance is estimated using the Permutation Jensen-Shannon Distance (PJSD), \textit{i.e.} the distance between the probability distribution of permutation patterns \cite{bandt2002permutation, leyva202220,zunino2022permutation} observed in the two time series. Recently, it has been shown that a wide range of complex phenomena can be scrutinised with this ordinal symbolic quantitative metric\cite{zunino2022permutation}. Taking into account the robustness to noise effects and the invariance under scaling of the data associated with the ordinal symbolisation approach \cite{leyva202220}, together with numerous applications of the corresponding distribution~\cite{zanin2008forbidden,martinez2018detection,mccullough2015time,mccullough2016counting,mccullough2017regenerating,bariviera2018analysis,sakellariou2016counting,kulp2016using,kulp2014discriminating,quintero2018differentiating,yamashita2022assessing,barreiro2011inferring,olivares2020contrasting,sun2014characterizing}, the PJSD arises as a quantitative metric especially suited for the analysis of real-world signals. We also consider this approach to be versatile because the hypothesis to be tested can easily be set up by conveniently choosing the time series taken as reference.  

The proposed model is tested using data about real landing operations in ten major European airports between years 2018 and 2020. We firstly describe the synthetic model, and explore how the resulting dynamics is changed by its defining parameters. We then recover the best parameters describing the observed dynamics, and show how the resulting synthetic time series retain great part of the original information. We further study how the correlation between consecutive values in the high-frequency term of the dynamics changes between airports. This represents how the landing of one aircraft depends on what happened in the previous hours, or, in other words, the memory term in the landing dynamics; and can thus be used as a metric of airport efficiency in handling arrival operations. We finally compare the dynamics pre and post COVID-19, and show how this has changed beyond what can be attributed to a simple reduction of traffic. We conclude by drawing some conclusions on the operational values of these findings.

\section{Ordinal pattern probability distribution}

Ordinal patterns are defined by two parameters: the order of the permutation symbols $D \geqslant 2$ ($D \in {\Bbb N}$, the pattern length) and the lag $\tau$ ($\tau \in {\Bbb N}$, the time separation between the values). Any one-dimensional time series, $X(t) = \{x_t ; t = 1, \dots,M\}$, can be mapped into subsets of length $D$ of consecutive ($\tau = 1$) or non-consecutive ($\tau > 1$) values, generated by $(t) \equiv (x_t, x_{t+\tau}, ..., x_{t+(D-2)\tau}, x_{t+(D-1)\tau})$, which assigns to each time $t$ the $D$-dimensional vector of values at times $t$, $t + \tau , .... , t + (D-1)\tau$. Each element of the vector is subsequently replaced by a number related to its relative ranking, {\it i.e.} the smallest value by zero and the largest one by $D-1$. By the ordinal pattern corresponding to the time ($t$) we therefore mean the permutation $\pi_{i}$ of $0, 1, . . . , D - 1$, representing the relative amplitude (strength) of each element in the original vector. Whenever values $X(t)$ are drawn from continuous distribution, equal values are very unusual~\cite{cuesta2018patterns}, digitised data with low resolution or discrete measurements could nevertheless have considerable number of equalities. In those cases, the most used recipe is to break these ties by adding a small amount of noise~\cite{bandt2002permutation}.

By counting the number of times each ordinal pattern $\pi_{i}$ appears in the symbolised time series $X(t)$, normalised by the total number of ordinal patterns $M-(D-1)\tau$, we can compute an ordinal pattern probability distribution function
\begin{equation}\label{Eq:PDF} 
    p_{i} = \frac{ \# (\pi_{i})}{M-(D-1)\tau}, \,\,\, i = 1,2, ...,D!,
\end{equation}
where $\#(\pi_{i})$ stands for the cardinality of $\pi_{i}$. To illustrate the ordinal pattern probability reconstruction process, we consider the simple case of a time series with ten ($M = 10$) values $X = \{1,5,7,9,6,10,0,12,1,15\}$, pattern length $D = 3$, and lag $\tau=1$, which gives 8 possible ordinal patterns. The first two triplets, (1,5,7) and (5,7,9), are mapped to the pattern (012), since the values are already sorted in ascending order. On the other hand, (7,9,6) and (6,10,0) correspond to the pattern (120), since $x_{t+2} < x_t < x_{t+1}$; while (9,6,10), (10,0,12) and (12,1,15) are mapped to the ordinal pattern (102) since $x_{t+1} < x_t < x_{t+2}$. Lastly, the triplet (0,12,1) maps to the pattern (021). The associated probabilities of the six ordinal patterns are therefore $p(012) = p(120) = 1/4$, $p(102) = 3/8$, $p(021) = 1/8$, and $p(201) = p(210) = 0$. 

Taking into account that there are $D!$ possible permutations of a $D$-dimensional vector, the condition $M \gg D!$ must be satisfied in order to obtain a reliable statistics~\cite{bandt2002permutation}; consequently, the maximum pattern length is limited by the length of the time series under analysis. Although in their seminal paper\cite{bandt2002permutation} Bandt and Pompe suggested to fix $\tau=1$, it has been demonstrated that the analysis with lagged data points, \emph{i.e.} $\tau \ge 2$, may be useful for reaching a better comprehension of the underlying dynamics of some systems, as it physically corresponds to multiples of the sampling time $\delta t$ \cite{zunino2010permutation,soriano2011distinguishing,zunino2012distinguishing,olivares2020multiscale}. In the present analysis we set $D=3$, as it gives a clearer graphical observation of the hierarchy of the ordinal patterns probability distribution, and due to the limited length of the time series under analysis; $\tau>1$ for representing the low-frequency daily dynamics of the landing {\color{red}flow}, and specifically, $\tau = 1$, for quantifying short-term correlations.

\subsection*{Permutation Jensen-Shannon distance}

The PJSD was recently introduced as a versatile metric to measure the degree of similarity between the symbolic ordinal sequence statistics of two time series~\cite{zunino2022permutation}. It is defined in terms of the Jensen-Shannon divergence~\cite{lin1991divergence} between two ordinal pattern probability distributions $P = \{p_1, ..., p_N\}$ and $Q = \{q_1, ..., q_N\}$ associated with the two time series under analysis,
\begin{equation}\label{eq1}
    D_{\text{JS}}(P,Q) = S((P+Q)/2) - S(P)/2 - S(Q)/2,
\end{equation}
where $S(P)= \sum^{N}_{i=1} p_i \ln p_i $, is the classical Shannon entropy. The PJSD is hence obtained by calculating the square root of Eq. \ref{eq1}, and its normalised version reads
\begin{equation}\label{eq2}
    \text{PJSD}(P,Q) = \sqrt{\frac{D_{JS}(P,Q)}{\ln 2}}.
\end{equation}
Since it constitutes a measure of distinguishability between two probability distributions, the ordinal patterns composition between different sequences can be quantitatively compared through this metric. Obviously, larger values of this quantifier indicate less similarity between the symbolic mapping of the signals, and vice versa. It is intuitively expected that signals coming from the same underlying dynamics would have small estimated values of this ordinal distance (close to but not exactly zero as a consequence of finite-size effects), whereas larger values (significantly different from zero) will be found when the signals have different dynamics. Actually, in the former instance, the PJSD has been shown to asymptotically converge to zero with the series size~\cite{zunino2022permutation}.

\section{Landing flow data}

We consider the landing activity from 10 major European airports in two different periods: from June 1st,
2018 to June 1st, 2019; and from June 1st, 2020 to June 1st, 2021. Table \ref{Tab_1} lists the airports considered in this work classified by their number of runways (one or two) dedicated to landings operations. Additionally, the second column reports their four-letter code defined by the International Civil Aviation Organization (ICAO). Hereinafter airports will be identified by this code.  

\begin{table}[h!]
	\centering
	\caption{List of airports considered in this study. First column: Airports location. Second column: 4-digit ICAO identification code. Third column: median length of the inactivity daily interval, respectively pre and post COVID-19 pandemic. Fourth column: median number of landings, pre and post COVID-19 pandemic.}
    	\begin{tabular}{l| c | c| c}
     		Airports &  ICAO & Inac. (hours) & $\#$landings \\ 
    		\hline \hline
    		one runway \\
    		\hline \hline
            Brussels	    & EBBR      & 3 / 3   & 13 / 4\\		 
            Dusselforf	    & EDDL      & 5 / 6   & 14 / 3\\
            Munich		    & 	EDDM    & 4 / 5   & 29 / 3\\
            London-Heathrow & 	EGLL    & 5 / 5   & 26 / 11 \\
            Milan-Malpensa  & 	LIMC    & 2 / 2   & 13 / 5 \\
            Vienna	    	& 	LOWW    & 3 / 4   & 22 / 5 \\
            Zurich	    	& 	LSZH    & 2 / 6   & 27 / 6\\
            \hline \hline
            Two runways  \\
            \hline \hline
            Amsterdam		        & 	EHAM    & 2 / 3 & 32 / 8 \\
            Frankfurt del Meno	    & 	EDDF    & 4 / 5 & 38 / 12 \\
            Madrid-Barajas		    & 	LEMD    & 2 / 3 & 22 / 8 \\
            
 			\hline     
       \end{tabular}
       \label{Tab_1}
\end{table}

\begin{figure}
\includegraphics[width=0.5\textwidth]{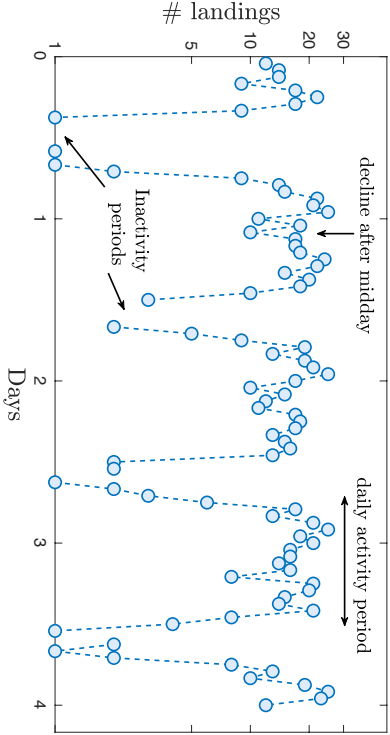}
\caption{\label{fig_vol_milan} 4-days evolution of the number of landings over a temporal window $\delta t=$ 1 hour, for Milan-Malpensa (LIMC) airport. This dynamic is representative for all landing flows considered in this work. }
\end{figure}

The time of landing of all aircraft arriving at each airport has been extracted from ADS-B position reports, obtained from the OpenSky Network (https://opensky-network.org)~\cite{schafer2014bringing}. ADS-B (Automatic Dependent Surveillance - Broadcast) is a technology allowing aircraft to continuously send radio messages, stating their position and other information of relevance~\cite{williams2009gps,salcido2017analysis}; these messages are then received by ground stations, and integrated into coherent reports. For each airport, flights in their final approach have been identified when the last known position was within a radius of 3 nautical miles from the center of the airport, and the last reported altitude below 500 meters above ground level; the landing (touch down) moment has then been obtained by modelling the last seconds of the flight. Note that this is an estimation of the landing time, and therefore that observational noise exists in the result. Once all landings have been identified for each airport, these are transformed into a time series of landing flows, by simply calculating the number of landings in temporal windows of $\delta t$= 1 hour. As a representative example, in Fig. \ref{fig_vol_milan} we show the landing flow in a 4 days period for Milan-Malpensa airport (LIMC). At a first view, the daily activity period is clearly identified, followed by inactivity intervals for which no landings are observed (\textit{i.e.} corresponding to nights). Moreover, a secondary long-term component during the activity time can be observed, \textit{i.e.} a decreased number of landings after midday.

Inactivity periods are not fixed, and in fact fluctuate. The third column of Tab. \ref{Tab_1} lists the median of the inactivity intervals in hours during the pre and post COVID-19 periods. Similarly, fourth column lists the median of the number of landing operations per hour.  

\begin{figure}[ht!]
\includegraphics[width=0.45\textwidth]{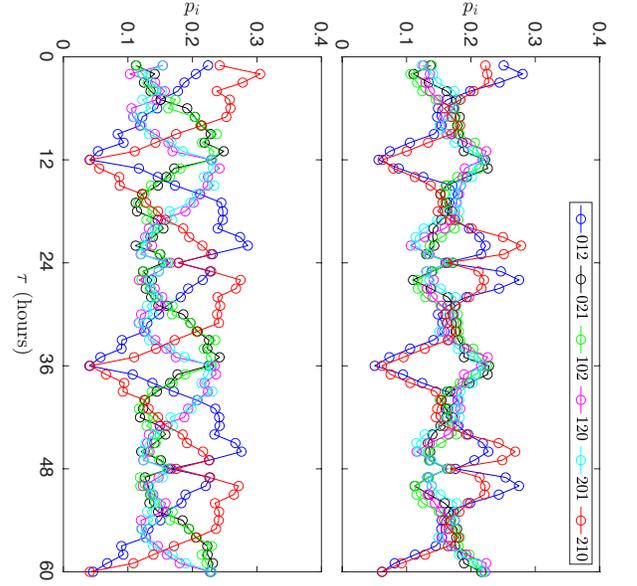}
\caption{\label{fig_exp_OPP} Ordinal pattern probabilities as a function of the lag $\tau$ for LIMC (top panel) and EHAM (bottom panel). Probabilities estimated from the annual period between June 1st, 2018 and June 1st, 2019. }
\end{figure}

We further report in Fig. \ref{fig_exp_OPP} the ordinal probability distributions as a function of $\tau$ for LIMC and EHAM airports, being representative of airports with respectively one and two runways dedicated to landings. We want to stress that, hereinafter, a small amount of noise has been added to break the ties present in the sequences. As expected, in both cases the probabilities oscillate regularly with a 24 hours periodicity. Note that the six ordinal patterns equally probable for $\tau=24, 48, ...$ since the added random perturbation breaks the equal amplitude values originally obtained when the lag matches the oscillation period and its multiples. This feature has been observed in oscillations with additive noise~\cite{carpi2018persistence} and stochastic oscillations~\cite{zunino2012distinguishing}. In the case of LIMC we observe that ordinal probabilities tend to cluster into two groups for almost all values of $\tau$, as opposed to EHAM, for which a different clustered structure is observed. It is important to stress that these results are representative for all the airports considered in this work.

\section{Landing flow as a Markov-modulated model}

\subsection*{Modulated Markov Jitter}

Aiming to represent sunspot activity, Blanter et al. \cite{blanter2006short} introduced a simple model of a modulated noise, more specifically a modulated Markov jitter, which is a high-frequency Markov signal multiplied by a long-term component. The sequence $\psi(t)$ is mathematically defined as:
\begin{equation}
    \psi(t) = F(t) \eta(t),
\end{equation}
where the function $F(t)$ represents any kind of long-frequency dynamics, while the high-frequency term $\eta(t)$ is an auto-regressive process of first order AR(1). The later term is thus defined as:
\begin{equation}\label{markov}
    \eta(t_{i+1}) = \alpha  \eta(t_{i}) + \xi(t_{i+1}),
\end{equation}
where $\xi(t)$ is an independent random variable uniformly distributed on the interval $[0,1]$, and the parameter $\alpha$ ($\alpha \in [0,1)$) characterises the correlation between two consecutive values. Note that for $\alpha=0$, $\eta(t)$ is equal to $\xi(t)$ and is a totally uncorrelated variable. For a sampling interval $\delta t = t_{i+1}-t_i$, the mean life-time of the process (\ref{markov}) is defined as\cite{crouzeix2006long}:
\begin{equation}\label{eqtheta}
    \theta = \frac{\delta t}{1-\alpha}.
\end{equation}

\subsection*{Simulating landing flow}

In order to simulate daily landing flow dynamics, we consider the two basic low-frequency regularities previously identified in the real data. These long-term components are modeled as:
\begin{equation}
    F(t;\omega, \phi,c) = \sin(\omega\,\, t_i + \gamma \zeta_i + \phi) + c.
\end{equation}
$\zeta_i$ is an independent random variable uniformly distributed on the interval $[0,1]$, with $\gamma$ defining its weight and thus controlling the skewness of the activity period. Additionally, $\phi$ is the phase; and non-negative values are guaranteed with $c \geq 1$. The parameter $\omega$ is used to introduce the observed periodicities in the reconstructed dynamics. We therefore consider a first daily regularity, representing the overall envelope of the landing volumes, and obtained for $\omega_1=2\pi/T$; note that $T=24-\nu$ hours, where $\nu$ is the night inactivity period for which $F(t,\omega)=0$---see Tab. \ref{Tab_1} for real values.
The second daily regularity encodes the reduction of activity observed around midday, and is obtained for $\omega_2 = 2\pi/(T/2)$.

It follows that a simple model of the landing flow fluctuations can be defined as:

\begin{equation}
    V(t) = \eta(t) [F(t;\omega_1,\phi_1,c_1) + \beta F(t;\omega_2,\phi_2,c_2) ],
\end{equation}
\noindent where the parameter $\beta$ controls the decreasing number of landings at midday, and $\eta(t)$ is the auto-regressive process AR(1) given by Eq. \ref{markov}. Table \ref{Tab_2} summarises the role of each parameter, and the meaning and value of all constants.

\begin{figure*}
\includegraphics[width=\textwidth]{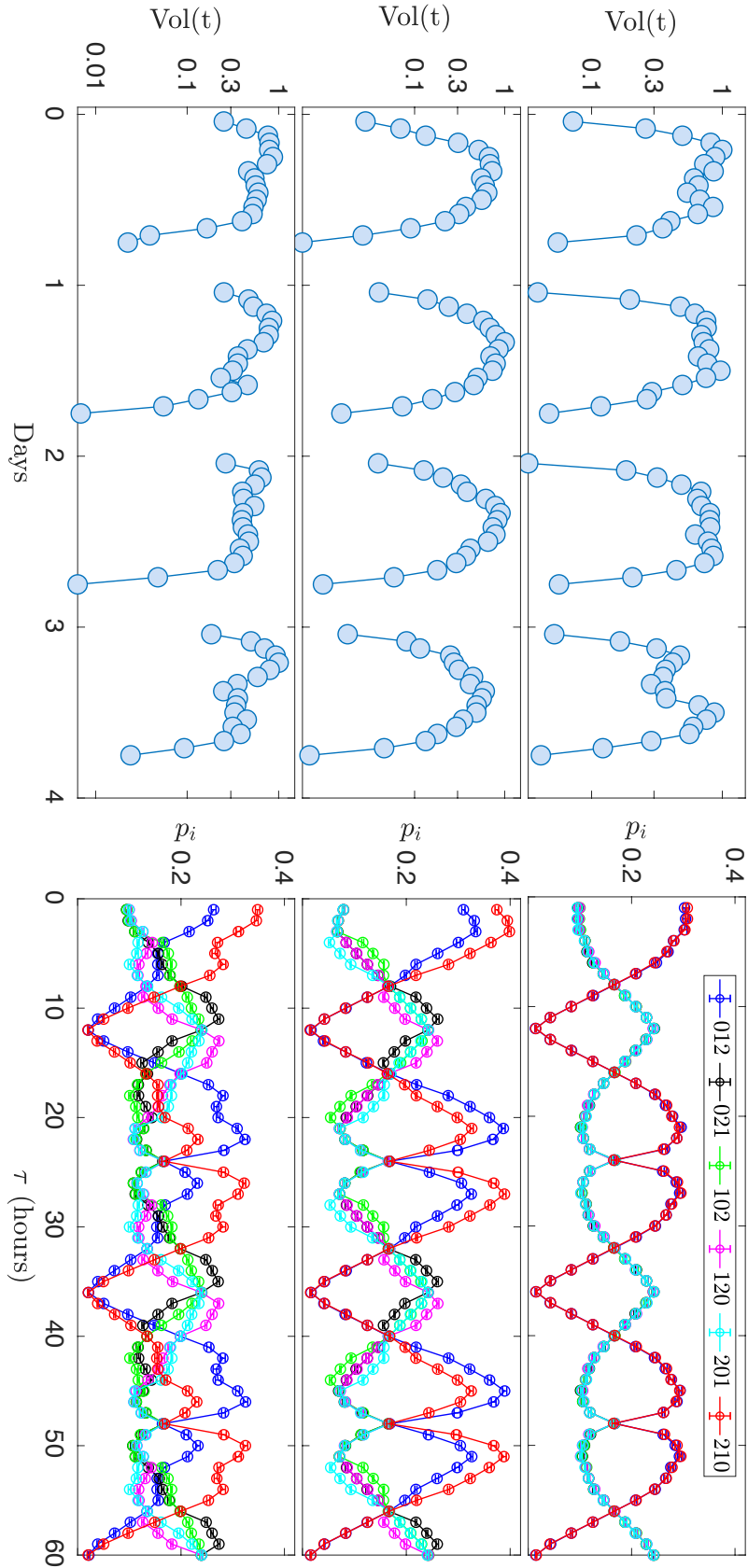}
\caption{\label{fig_model_OPP} (Left panels) 4-Days evolution of the normalized Markov-modulated model $V(t)$ for $\alpha=0.7$ with $\beta=0.5$ and $\gamma=0$ (top left panel), $\beta=0$ and $\gamma=0.5$ (middle left panel) and $\beta=1$ and $\gamma=1$ (bottom left panel). (Right panels) Corresponding ordinal pattern probabilities as a function of $\tau$, for synthetic time series representing one year of activity. Points represent the average of one hundred independent realisations, and error bars indicate one standard deviation. $\nu = 6$ for all cases.}
\end{figure*}

The left panels in Fig.  \ref{fig_model_OPP} show one realisation of the model for three different ensembles of the parameter triplet $\{\alpha;\beta;\gamma\}$, with a sampling window of one hour, and an activity period of $T=18$ hours, \textit{i.e.} $\nu=6$. Note that $\alpha$ is kept constant at $0.7$ to better visualise the effects of the other two parameters on the modulation. In all cases, the dominating low-frequency component is the 24 hours period (18 hours of activity plus 6 hours of no landing). For the triplet $\{0.7;0.5;0\}$ (top left panel), a symmetric daily modulation is obtained, since no stochastic component is added to the phase ($\gamma=0$); and a small decreasing of landings is appreciated around midday ($\beta=0.5$). In the case of $\{0.7;0;0.5\}$ (middle left panel), the symmetry between the opening and closing times of activities is slightly broken, and no decrease in activity is observed at midday as it is expected since $\beta=0$. Finally, for the triplet $\{0.7;1;1\}$ (bottom left panel), the broken symmetry and the midday modulation are even more evident. By comparing these synthetic time series with the real landing flow of LIMC in Fig. \ref{fig_vol_milan}, we conjecture that, by appropriately tuning the parameter triplet, a simple and adequate qualitative modelling of the landing flows can be achieved. 

\begin{table}[h!]
	\centering
	\caption{Summary of the parameters and constants of the Markov-modulated model.}
    	\begin{tabular}{c|c|c }
    	    \hline \hline
     		Parameters &  Quantifies & value  \\ 
    		\hline
            $\alpha$	& High-frequency correlation & -   \\
            $\beta$	    & Decrease of landings after midday &  -  \\
            $\gamma$    & Skewness of activity period & -\\
            \hline \hline
            Constants & Meaning & value \\
            \hline \hline
            $\nu$       & Duration of the inactivity interval  & see Tab. \ref{Tab_1} \\
            $T$         & Period of the activity oscillation & 24-$\nu$ \\
            $\omega_1$  & Activity oscillation frequency & $2\pi/T$ \\
            $\omega_2$  & Midday oscillation frequency & $2\pi/(T/2)$ \\
            $\phi_1$    & Activity oscillation phase & $\pi/2$ \\
            $\phi_2$    & Midday oscillation phase & $\pi/2$ \\
            $c_1$       & Guarantees positive values & 1 \\
            $c_2$       & Guarantees positive values & 1 \\
 			\hline     
       \end{tabular}
       \label{Tab_2}
\end{table}

Low-frequency components can be characterised by analysing the evolution of the ordinal pattern probability distribution as a function of $\tau$. The right panels of Fig. \ref{fig_model_OPP} depict the evolution of the ordinal patterns probabilities of the corresponding model realisations shown on the left. In all the cases, the probabilities oscillate regularly with a period of 24 hours.  More specifically, for the triplet $\{0.7;0.5;0\}$, we obtain a daily oscillation of the probabilities (top right panel), with a clustering of ordinal patterns into two groups for all temporal scales, \textit{i.e.} $p_{123}=p_{321}$ and $p_{132}=p_{213}=p_{231}=p_{312}$. This clustered structure has been theoretically validated for Gaussian processes with stationary increments~\cite{bandt2007order}, and also numerically and experimentally observed in noisy oscillations~\cite{carpi2018persistence} and real-world fluctuations~\cite{olivares2016quantifying}. On the other hand, for the case of $\{0.7;0;0.5\}$ and $\{0.7;1;1\}$ (middle and bottom right panels, respectively) these two pattern probability clusters are broken. When compared with the hierarchical ordinal structures as a function of $\tau$ obtained for the real landing flows (Fig. \ref{fig_exp_OPP}), they present enough similarities to confirm that a good approximation of the ordinal pattern probability distribution of the long-term oscillations of the real landing flows can be obtained.

\subsection*{Optimising the model parameters}

Once the model has been defined, the next logical step involves tuning it to retrieve the dynamics observed in each airport. We here propose to make use of the PJSD to estimate the set of parameters that better reproduce the ordinal hierarchy of real landing flows. This involves obtaining the set of parameters that minimise the PJSD between the model and the real data, with $\tau=1$, \textit{i.e.} defining an ensemble of parameter estimators $\{\Hat{\alpha},\Hat{\beta},\Hat{\gamma} \}$, as the set of values that minimise the PJSD:

\begin{equation}
   \{ \Hat{\alpha},\Hat{\beta},\Hat{\gamma} \} = \{ \alpha,\beta,\gamma : \text{PJSD}(P_M,Q_C) \,\, \text{is minimum}\},
\end{equation}
\noindent where $P_M$ and $Q_C$ are the ordinal pattern distributions for the model and the control sequence, respectively.

\begin{figure}[h!]
\includegraphics[width=0.45\textwidth]{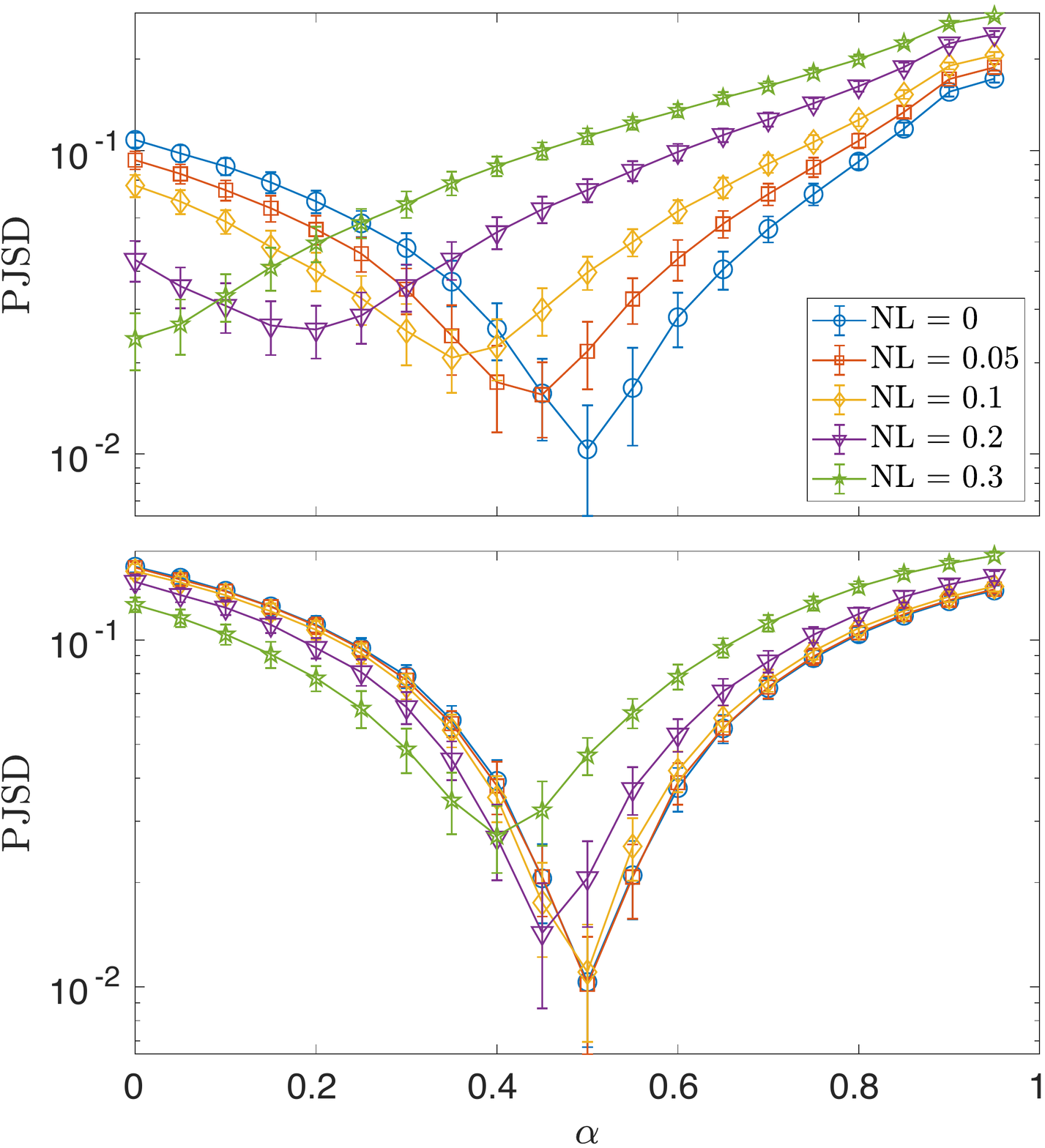}
\caption{\label{fig_noise_eff} PJSD between one control realisation with $\alpha_c=0.5$, $\beta_c=0.8$ and $\gamma_c=0.5$, and the model (with $\beta_c$, $\gamma_c$ and $\alpha$ free), using the fluctuation (top panel) and its integration (bottom panel). The noise level, NL, is defined as a multiple of the standard deviation of the original added Markov noise. Data points correspond to the average over one hundred independent realisations, and error bars to one standard deviation. $\nu = 6$ for all cases.}
\end{figure}

It has previously been shown that, from an ordinal pattern point of view, temporal correlations in fractional Gaussian noises are better quantified from the integrated fluctuations rather than using the fluctuations themselves~\cite{olivares2016quantifying,zunino2022permutation}, especially when crossovers between different scaling regimes are present~\cite{olivares2020multiscale}. It should be added here that for integration of a sequence $X(t)$ we mean its cumulative sum:   $\sum_t^M (x_t -\langle X \rangle)$. With this in mind, we have compared the minimisation procedure of the PJSD when using the modulated fluctuation and its integration. To keep the comparison simple, we fix the parameters $\gamma=0.5$ and $\beta = 0.8$, and minimise the PJSD between a control sequence characterised by $\alpha_c = 0.5$ and the Markov-modulated model with $\alpha$ free. Observational noise has been added to the sequences to simulate the landing time uncertainty. These results are depicted in Fig. \ref{fig_noise_eff}. We found that, when noisy fluctuation are considered, the value of $\alpha$ for which the PJSD reached a minimum is significantly smaller than $\alpha_c$ (see top panel in Fig. \ref{fig_noise_eff}). This contrasts with the results obtained for the integrated sequences, for which a more robust identification of the temporal correlation is confirmed (see bottom panel in Fig. \ref{fig_noise_eff}). This analysis leads to conclude that the minimization procedure applied on the integrated fluctuations allows to obtain a better high-frequency optimization of the Markov-modulated model especially within noisy environments.

\begin{figure}[h]
\includegraphics[width=0.5\textwidth]{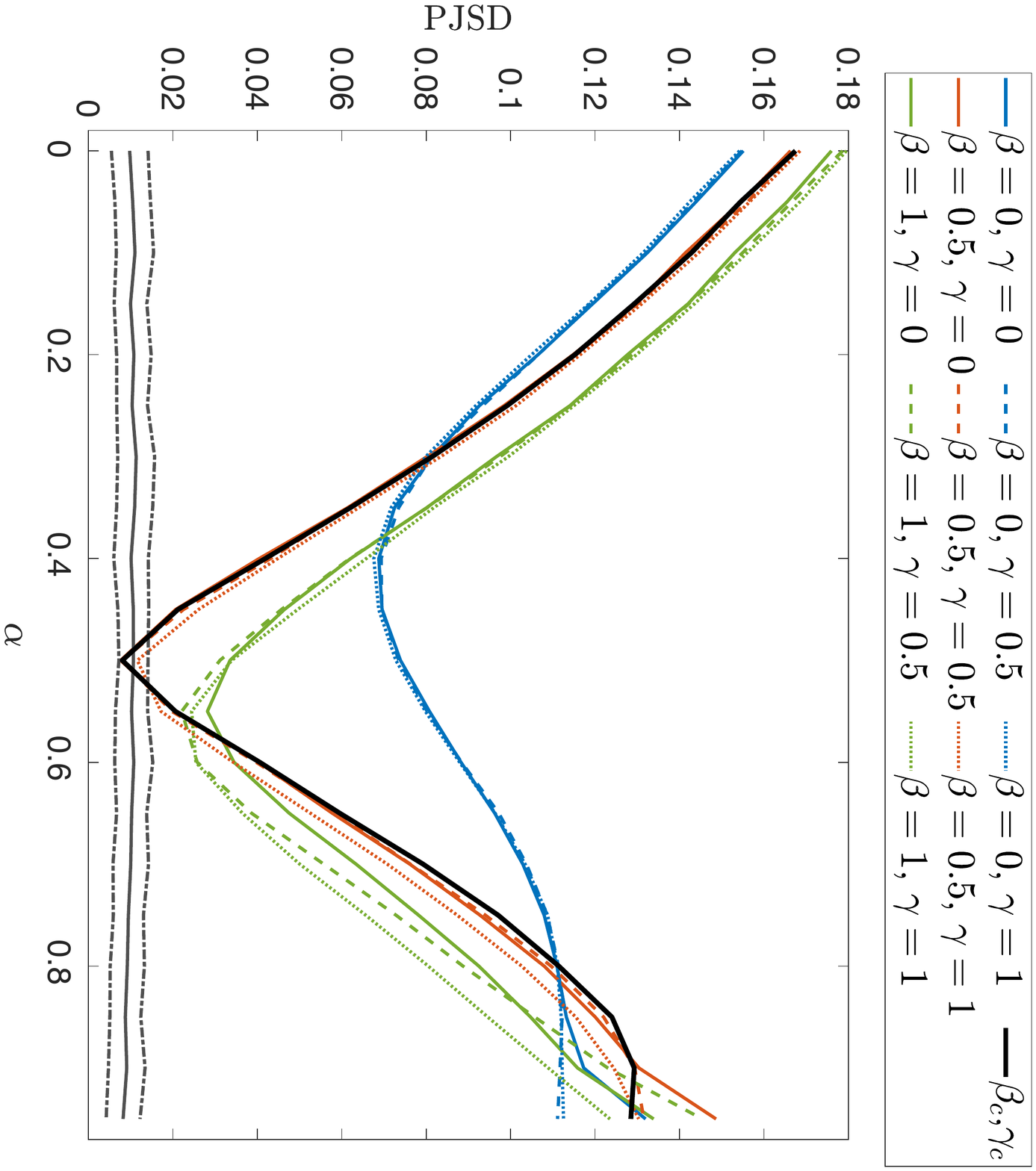}
\caption{\label{fig_JSD_3par} PJSD between the Markov-modulated model with free parameters, and one control realisation with $\alpha_c=0.5$, $\beta_c=0.5$, $\gamma_c = 0.3$, and $\nu = 6$. The analysis is performed on the integrated fluctuations and the average over one hundred independent realisations is plotted. Solid and dashed grey lines indicate, respectively, the mean and $\pm$ one standard deviation of the baseline obtained by estimating the PJSD between two independent realisations of the model.}
\end{figure}

To further illustrate the tuning procedure, we perform a minimisation of the PJSD between a control sequence obtained with the model, with $\alpha_c =\beta_c=0.5$ and $\gamma_c = 0.3$; and sequences generated from the proposed model with the following ranges of parameters $0\leq \beta \leq 1$, $0\leq \gamma \leq 1$ and $0\leq \alpha <1$. Taking into account the results obtained previously, this optimisation analysis is performed directly on the integrated fluctuations. The evolution of the PJSD as a function of $\alpha$ for different values of the modulation parameters is depicted in Fig. \ref{fig_JSD_3par}. For the sake of clarity, we have depicted results for some selected pairs of values of $\beta$ and $\gamma$. Moreover, a baseline, that take into account finite-size and amplitude distribution effects~\cite{zunino2022permutation}, is established by estimating the PJSD between two independent realisations of the model. On one hand, we found that $\beta$ strongly affects the minimisation, as opposed to $\gamma$, which has a much less noticeable effect. This is because $\gamma$ only quantifies the stochastic fluctuation of the intra-day phase, with a minimal impact on the associated ordinal pattern probability distributions at $\tau=1$. On the other hand, a global minimum is obtained for values of $\alpha$ and $\beta$ near the control parameters $\alpha_c$ and $\beta_c$ (see solid black line). More specifically, the global minimum happens at $\Hat{\alpha} = 0.51 \pm 0.02$,  $\Hat{\beta} = 0.52 \pm 0.04$ and $\Hat{\gamma} = 0.34 \pm 0.23$. The large standard deviation of $\Hat{\gamma}$ suggests that its estimation becomes irrelevant in this context. The minimum distance reaches the baseline range, indicating that the ordinal pattern probability distribution of the control sequence can accurately be reproduced by the proposed model.

\section{Quantifying short-range correlations in landing flows}

Moving to a deeper analysis, the real landing data have been used to tune the parameters $\alpha$ and $\beta$ of the model; a different instance of the model has been considered for each airport, in order to recover the ordinal pattern probability distribution of the hourly evolution of the landing flow for each of them. Note that we use $d = 3$ and $\tau=1$ in this analysis, thus each ordinal pattern encodes temporal information about the landing flow throughout $3$ hours. 

\begin{figure}
\includegraphics[width=0.5\textwidth]{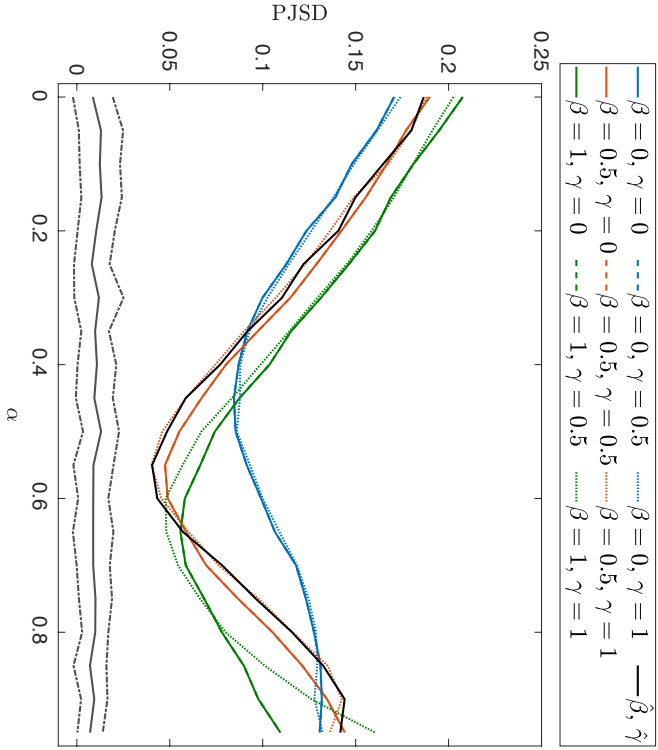}
\caption{\label{fig_JSD_ex} PJSD between the integrated realisation of the Markov-modulated model ($\nu=2$ hours) and the integrated hourly landing flow of the LIMC airport. Solid and dashed grey lines indicate, respectively, the mean and $\pm$ one standard deviation of the baseline obtained by estimating the PJSD between two independent realisations of the model. Average over one hundred independent realisations is plotted. }
\end{figure}

\begin{figure}
\includegraphics[width=0.48\textwidth]{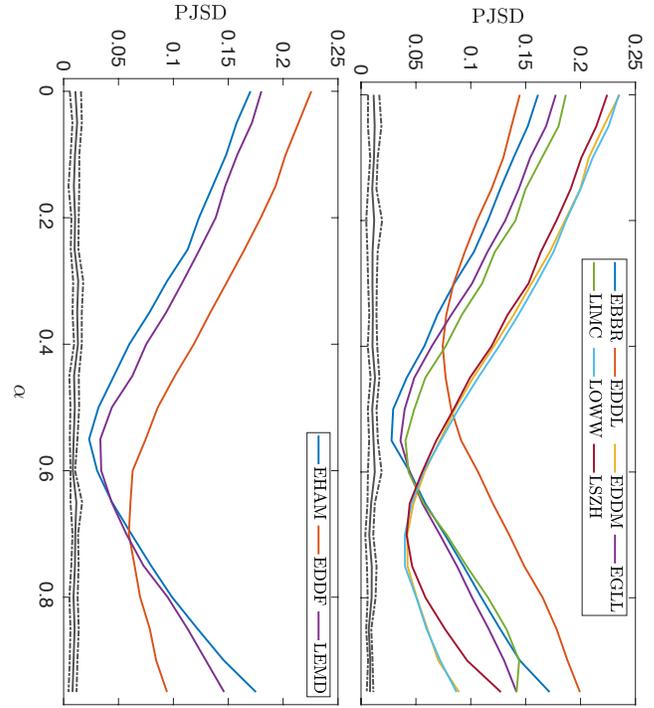}
\caption{\label{fig_minJSD_all} PJSD between integrated realisations of the Markov-modulated model (for $\nu$ see Tab. \ref{Tab_1}) and the integrated landing flow for all airports. Solid and dashed grey lines indicate, respectively, the mean and $\pm$ one standard deviation of the baseline obtained by estimating the PJSD between two independent realisations of the model. Average over one hundred independent realisations is plotted. }
\end{figure}

We firstly consider a complete year of operations, spanning between June 1st, 2018 and June 1st, 2019. The evolution of the PJSD between the integrated real and synthetic time series has been obtained, varying the parameters $0 \leq \alpha < 0.95$, $0\leq \beta \leq 1$ and $-1 \leq \gamma \leq 1$, with a step of $0.05$. Additional constants include $c_1 = c_2=1$, $\phi_1=\phi_2= \pi/2$, $\delta t = 1$ hour, and $\nu$ being equal to the median of the inactivity intervals as per Tab. \ref{Tab_1}---see also Tab. \ref{Tab_2} for a synthesis. As an example, Fig. \ref{fig_JSD_ex} reports the result for LIMC. As it has previously been observed for control sequences, for a given value of $\beta$, the stochastic parameter $\gamma$ becomes irrelevant to estimate the temporal correlation, \textit{i.e.} the behaviour of the PJSD as a function of $\alpha$ is independent of $\gamma$ for practical purposes (at least for $D=3$ and $\tau=1$); its estimation has thus been discarded in what follows. The global minimum reached in this figure allows to estimate the optimal parameters for modelling the hourly landing flow of LIMC airport, namely $\Hat{\alpha}=0.55\pm 0.01$ and $\Hat{\beta} = 0.52 \pm 0.05$ (see solid black line in Fig. \ref{fig_JSD_ex}). As a summary, Fig. \ref{fig_minJSD_all} depicts the evolution of the PJSD as a function of $\alpha$ for all airports (all other parameters having already been optimised); for the sake of clarity, the top panel reports results for airports with one landing runway, and the bottom panel for two runways. Deviations from the baseline reference are found; this is not much surprising, considering that the present model is a simple one based on a modulated linear noise, and that a plethora of additional phenomena (\textit{e.g.} weather, technical failures, and so forth) impact on real operations. Moreover, the unavoidable presence of observational noise tends to move the PJSD value at the minimum upwards, as it was numerically shown in Fig. \ref{fig_noise_eff}.

\begin{figure}[h!]
\includegraphics[width=0.5\textwidth]{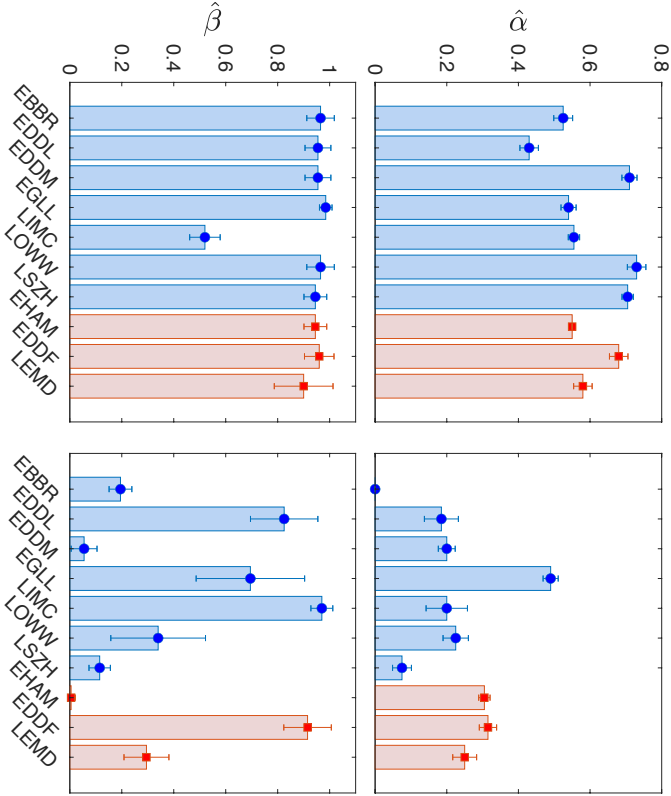}
\caption{\label{fig_par} Average estimated values $\Hat{\alpha}$ (top panels) and $\Hat{\beta}$ (bottom panels) for all airports considered in this study, for a one-year window pre- (left panels) and post- (right panels) COVID-19 pandemics. Blue bars correspond to airports with one landing runway, red ones to airports with two landing runways.}
\end{figure}

Top and bottom left panels in Fig. \ref{fig_par} respectively show the estimations of $\Hat{\alpha}$ and $\Hat{\beta}$ for each airport. We observe no significant differences between the results of airports with one and two runways. In all cases, we found a medium to high degree of correlation, being LOWW the airport with the largest $\Hat{\alpha}=0.73 \pm 0.02$ and EDDL the one with the smallest $\Hat{\alpha}=0.43 \pm 0.02$. On the other hand, estimations of $\Hat{\beta}$ are quite similar across airports, except for LIMC. 

In order to study the impact that the COVID-19 pandemic had on the dynamics of the landing flow, we have considered a second year window, spanning between June 1st, 2020 and June 1st, 2021. This thus includes the same months as the original one, taking seasonal effects into account and avoiding March and April 2020, which had a completely different dynamics and a major reduction in operations. The estimations of $\Hat{\alpha}$ and $\Hat{\beta}$ for this period are respectively shown in the top and bottom right panels of Fig. \ref{fig_par}. A significant decrease in the correlations is found in all cases; Brussels airport (EBBR) presents the most extreme one, with $\Hat{\alpha}$ dropping from $0.52$ to $0$. This implies a purely aleatory high-frequency dynamics. In the case of $\Hat{\beta}$, we estimate heterogeneous values, as opposed to the results obtained for the pre-pandemic period; these differences will further be discussed and explained below.

\subsection*{Comparing real and simulated data}

As common with any synthetic model, a natural question that ought to be tackled is what share of the original dynamics is retained by the synthetic time series; or, in other words, if the Markov-modulated model here used is enough to represent the specificities of landing dynamics. We here transform this question in a classification problem, in which a Deep Learning model \cite{lecun2015deep} is used to identify time series belonging to a pair of airports. Specifically, in each classification task, two sets of time series are used as input---one set for each considered airport. The model is then asked to predict the airport corresponding to a new time series; the fraction of time this prediction is correct (also known as the accuracy) is then averaged, and used as indicator of the classification score. The underlying idea is that, if the classification score between two airports remains high when using the synthetic time series, we can then conclude that what makes each airport unique (or its identifiability) is also preserved, and hence that the corresponding synthetic time series are a good representation of the real dynamics.

We resorted to a Residual Network (ResNet) Deep Learning model, \textit{i.e.} an artificial neural network in which connections between layers are not sequential, but is instead characterised by the presence of links between two distant layers without involving the intervening ones \cite{he2016deep, wang2017time}. These layers are organised in blocks, each one composed of a convolutional layer, followed by a batch normalisation and a rectifier linear unit. The final layer of each block is added to the input layer of the same. Finally, the output layer is again a fully connected layer with a softmax activation. This model has been chosen for its high performance in many real-world problems \cite{wang2017time, zanin2022can, ivanoska2022assessing}.

\begin{figure}[h!]
\includegraphics[width=0.45\textwidth]{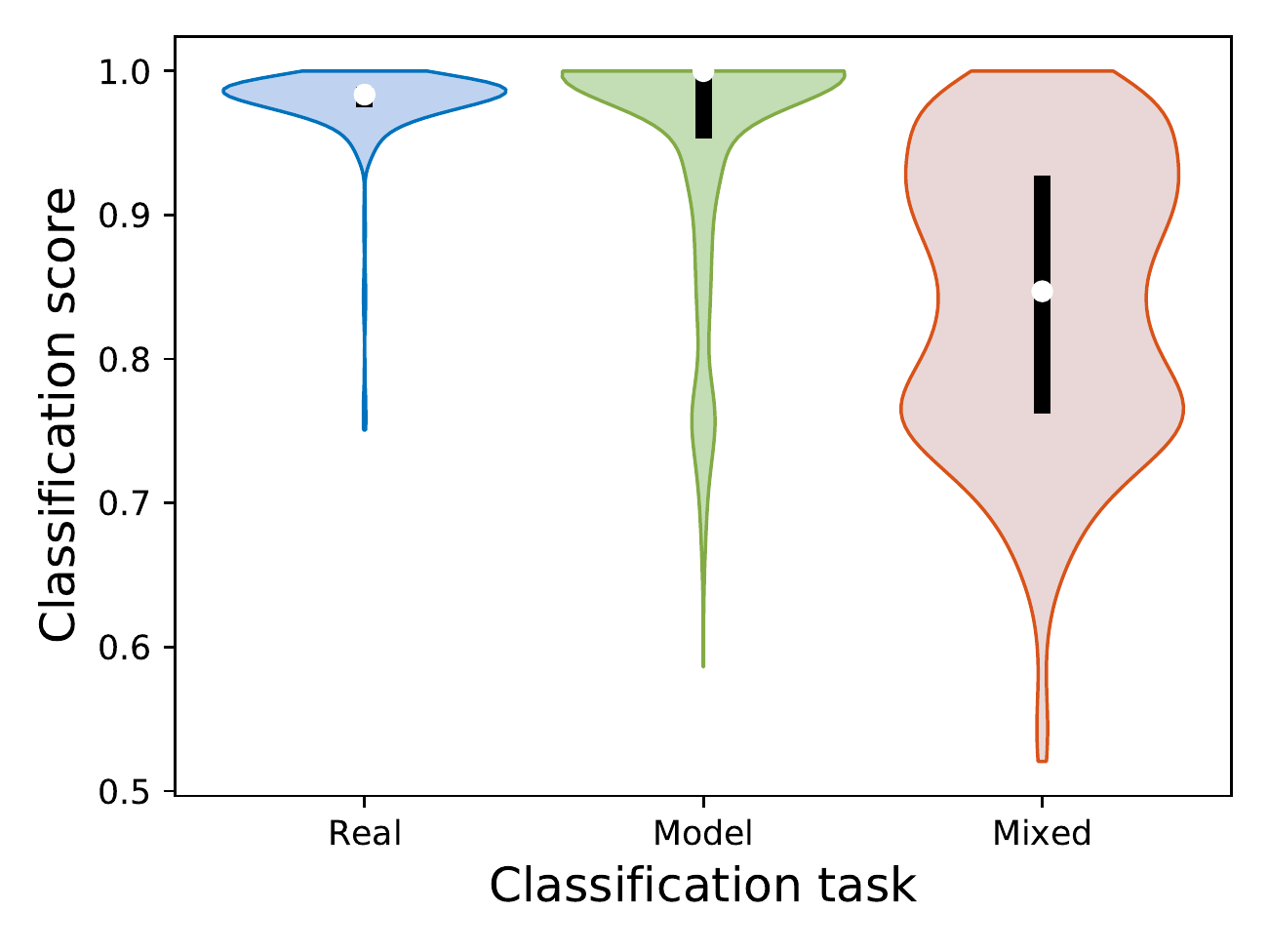}
\caption{\label{fig_DL} Distributions of the scores obtained in classification tasks executed over sets of time series representing the dynamics of pairs of airports, using a Residual Network Deep Learning model - see main text for details. From left to right, the three violin plots correspond to: the classification of the real time series; the classification of the time series yielded by the model; and tasks in which the model is trained with the real data, but validated using the synthetic ones. White dots represent the median of the distribution; black vertical bars the $25$--$75$ percentiles. }
\end{figure}

As a first analysis, the left blue violin plot of Fig. \ref{fig_DL} reports the distribution of the score obtained by classifying the real landing flow time series---values correspond to all airport pairs, and $10$ realisations of each classification task. It can be appreciated that the mode of the score (represented by the white circle) is very close to $1$, and thus that the ResNet model is able to correctly identify the source of the time series almost all the times. Secondly, the middle green violin plot represents the same distribution when only using synthetic time series. In this latter case, the score is even higher, suggesting that the model creates time series that are highly unique, hence distinguishable and identifiable, for each airport. Finally, the latter violin plot represents the score when the ResNet model is trained using the real time series, but then evaluated on the synthetic ones. While the median of the score substantially drops, it is still above $85\%$. This indicates that, while part of the information is lost, a great share of what makes each airport unique is retained in the synthetic time series; and thus that these still represent the essential elements of each airport's dynamics.

\subsection*{The influence of missing flights}

Due to the very nature of the ADS-B technology used to reconstruct landing events, flights may be missing from the data set ---as \textit{e.g.} a receiver may not be on-line all the time, may have a bad reception and can thus miss some position reports, or technical problems may prevent a correct integration of the information. Due to such potential incompleteness, we here propose  a numerical experiment to evaluate the impact of missing flights in the estimation of short-range correlations. 

\begin{figure}[h!]
\includegraphics[width=0.49\textwidth]{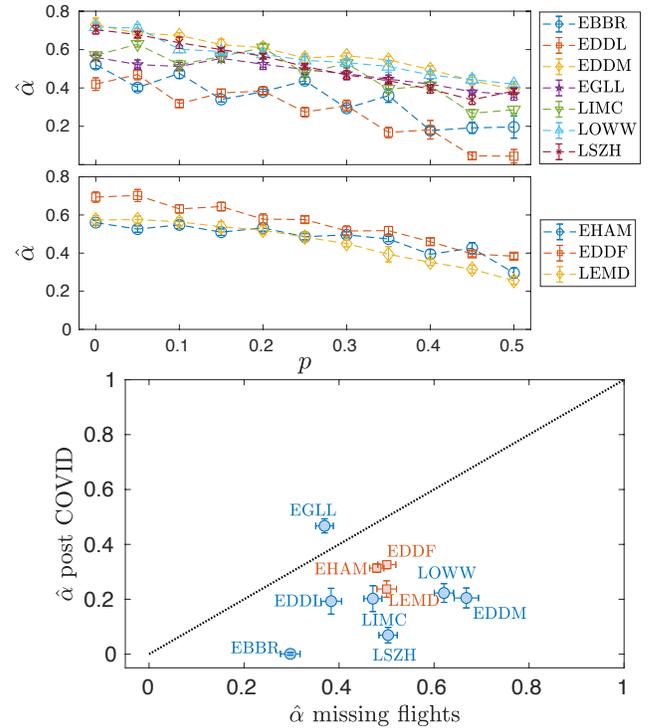}
\caption{\label{fig_missing} Correlation parameter $\Hat{\alpha}$ as a function of the percentage of missing flights for airports with one and two landing runways, top and middle panels respectively. Bottom panel: $\Hat{\alpha}$ estimated from the post-pandemic period versus $\Hat{\alpha}$ estimated in the pre-pandemic period but with adjusted traffic. Values correspond to the average of one hundred independent realisations, and error bars indicate one standard deviation.}
\end{figure}

Specifically, starting with the real time series corresponding to each airport, a fraction $p$ of landing events have been randomly deleted, for then estimating the new correlation $\Hat{\alpha}$. The evolution of the new $\Hat{\alpha}$ as a function of $p$ is shown in top and middle panels in Fig. \ref{fig_missing}. As may be expected, the former decreases with the latter, yet the decay of $\Hat{\alpha}$ is quite slow. Due to the sampling temporal window of $\delta t=1$ hour, a large $p$ is needed to break the ordinal ranking of the 3-hours patterns ($D=3$), and hence to significantly disrupt the dynamics. As might be expected, airports with less traffic volume show a slightly faster decay of the correlation, such as EDDL (represented by the red open square in top panel of Fig. \ref{fig_missing}). 

A natural question that could be posed involves the nature (and causes) of the change in dynamics seen after the COVID-19 pandemic, specifically in Fig. \ref{fig_par}. It is well known that traffic was substantially reduced even one year after the pandemics, mainly due to the restrictions enforced on international mobility. As seen above, this can result in a decrease of $\Hat{\alpha}$, yet other factors can also have contributed.

To answer this question we performed an additional numerical experiment, comparing the correlations after COVID-19 with those obtained from the pre-pandemic time series, with the latter ones being adjusted to include a similar number of operations; in other words, landing events in the pre-pandemic period have been randomly deleted to achieve the same traffic volume observed in the post-pandemic period. The bottom panel of Fig. \ref{fig_missing} depicts the scatter plot of $\Hat{\alpha}$ estimated in the post-pandemic period as a function of the estimated $\Hat{\alpha}$ for the pre-pandemic period with adjusted traffic. It is observed that the correlations after COVID-19 are considerably smaller.  This suggests that the increase of the randomness observed after COVID-19 is not only due to a reduction in traffic; or at least that the correlations reduced more than what would be expected if interactions between aircraft would have been maintained irrespectively of the traffic volume.

\subsection*{Mean life-time}

\begin{figure}
\includegraphics[width=0.49\textwidth]{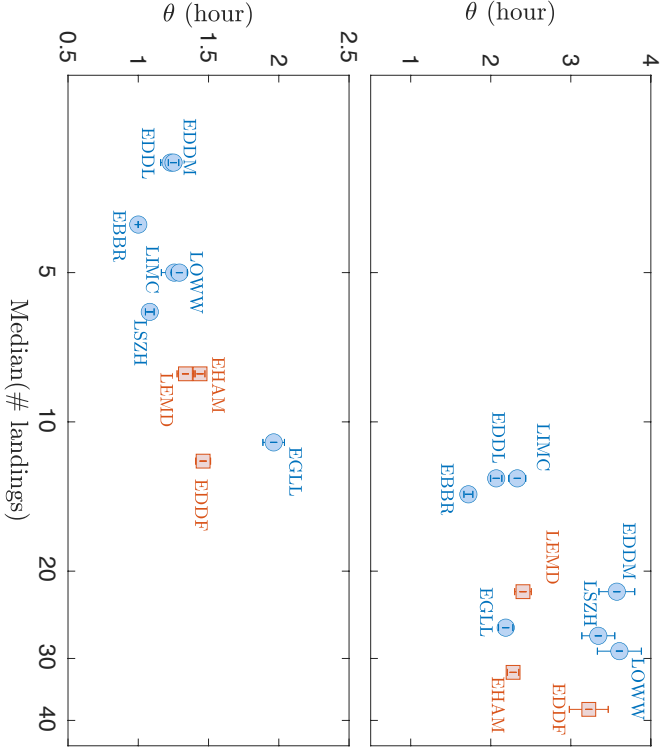}
\caption{\label{fig_lifetime} Mean life-time $\theta$ as a function of the median of the landing volume per hour, for pre and post COVID-19 periods, respectively top and bottom panels. }
\end{figure}

Lastly, one of the most relevant parameters that can be estimated is the mean life-time $\theta$. This can be interpreted as the mean duration of the dissipation of the airport landing events, or as how long do perturbations propagate in landing operations. This parameter can directly be estimated from the correlation parameter $\Hat{\alpha}$ by using Eq. \ref{eqtheta}. Figure \ref{fig_lifetime} shows the mean life-time as a function of the median of the hourly volume for all airports, considering both pre- (top panel) and post- (bottom panel) pandemic data. A non-trivial correlation between $\theta$ and the traffic volume can be observed. Additionally, the slope is larger (\textit{i.e.} $\theta$ increases faster with increasing traffic) in the pre-pandemic period; this suggests that airports were closer to a saturated dynamics, in which perturbations were difficult to dissipate and could hence propagate for a long time. The implication of these results will be discussed in the next section.

\section{Discussion and conclusions}

In this contribution we have presented the possibilities offered by the use of a Markov-modulated model to study the dynamics of landing events at an airport. Such model is extremely simple, and effectively only depends on two parameters ($\alpha$ and $\beta$). Additionally, the tuning of those parameters has been aimed at reproducing the structure of permutation patterns observed in the real data---as opposed to a minimisation of an amplitude-based distance metric. In spite of these simplifications, the time series synthesised by the model reproduce quite well the real dynamics, as demonstrated through a Deep Learning task---see Fig. \ref{fig_DL}. Most notably, the model can then be used to extract an estimation of the mean life-time $\theta$ for each airport, and to compare it in different conditions---\textit{e.g.} before and after the COVID-19 pandemic, as shown in Fig. \ref{fig_lifetime}.

The parameter $\theta$ represents the average duration of correlations between landing operations; or, in other words, the time scale over which landing aircraft can influence each other. This represents a well known dynamics in air traffic control; due to the limited capacity of airports, an external event (\textit{e.g.} adverse weather) may force a set of aircraft to delay their landing, in turn forcing later flights to also delay their landings in a cascade effect. In spite of the conceptual simplicity, the detection of these cascades is not a trivial task, as many other factors (\textit{e.g.} the capacity of nearby airspaces) may also intervene. This contribution shows how the cascade effect can be detected by relying on a tailored model and on statistical physics concepts; and additionally, how information about the micro-scale interactions between aircraft can be extracted from macro-scale observables.

Focusing on $\theta$, general correlation between this and the traffic volume is observed. This is to be expected, as landing aircraft can interact only in a congested environment, \textit{i.e.} when the availability of the landing runway is not guaranteed. The fact that such relation is not perfect is also to be expected, as many factors influence the capacity of a runway---as, for instance, the separation required to avoid wake vortex turbulences between consecutive landing aircraft depends on their relative size, and wind speed and direction. The comparison of the $\theta$ obtained for two different airports should thus be performed with due care. Still, it stands to reason to expect a non-linear behaviour of $\theta$ as a function of the traffic volume, for instance a supra-linear increase. This can easily be visualised by considering the extreme situation of traffic volumes approaching the maximum landing capacity: perturbation may propagate for an infinite amount of time, and hence $\theta \rightarrow \infty$.

Provided the capacity of a given airport remains constant, the proposed approach could be used as a metric for assessing the evolution of its efficiency, the latter understood as the capacity of handling a given volume of traffic while generating the minimum amount of interactions between aircraft. For instance, $\theta$ could be estimated in one-week windows; if the obtained value substantially deviates from what observed in other windows with a similar traffic volume, it can then be concluded that operations were disrupted, and that a deeper analysis may be required. It is yet important to highlight that a comparison between two $\theta$s can only be performed \textit{caeteris paribus}, as otherwise an increase in the value of the mean life-time of perturbations does not imply a reduction of efficiency \textit{stricto sensu}. To illustrate, one may conclude that the most efficient airport is one where no aircraft lands, such that no interactions can propagate and $\theta = 0$; while true from a numerical perspective, such airport would make no operational sense, and would not be economic efficient nor socially useful. 

Several additional aspects will have to be studied, before the idea here proposed could be deployed in an operational context. First of all, we have here considered one hour windows, as this is the standard to define the capacity of airports---in spite of alternative solutions having been proposed decades ago \cite{gilbo1993airport}. Other window lengths could nevertheless be tested; and it could even be hypothesised that such length should be tuned for each airport, to account for the different time separations at landing---due \textit{e.g.} to different baseline traffic volumes. Using shorter time windows would also increase the quantity of available data for each day, hence improving the tuning process. At the same time, it is not clear how the estimation of the parameters of the model is affected by the temporal resolution of the data; this could be relevant if a real-time solution is to be achieved. Finally, the whole solution must be validated, possibly relying on the feedback of expert users and airport managers.

In conclusion, while this contribution only proposes a first example, it also demonstrates the possibilities offered by an analysis of macro-scale aeronautical data, and the information about the micro-scale dynamics that statistical physics allows to extract from them.

\begin{acknowledgments}
This project has received funding from the European Research Council (ERC) under the European Union's Horizon 2020 research and innovation programme (grant agreement No 851255). M.Z. and F.O. acknowledges the
Spanish State Research Agency through the Mar\'ia de Maeztu project CEX2021-001164-M funded by the  MCIN/AEI/10.13039/501100011033. L.Z. gratefully acknowledge financial support from Consejo Nacional de Investigaciones Cient\'ificas y T\'ecnicas (CONICET), Argentina. 
\end{acknowledgments}

\bibliography{biblio}

\end{document}